\documentstyle[aps,multicol]{revtex}
\begin{document}

\draft
\title{\bf Current responses and voltage fluctuations\\
in Josephson-junction systems}
\author{Mahn-Soo Choi$^1$, M.Y. Choi$^2$, and Sung-Ik Lee$^1$}
\address{$^1$ Department of Physics, 
 Pohang University of Science and Technology, Pohang 790-784, Korea}
\address{$^2$ Department of Physics and Center for Theoretical Physics,
 Seoul National University, Seoul 151-742, Korea}

\date{To appear in Europhys.~Lett.}
\maketitle

\thispagestyle{empty}

\begin{abstract}
We consider arrays of Josephson junctions as well as
single junctions in both the classical and quantum-mechanical regimes,
and examine the generalized (frequency-dependent) resistance, which 
describes the dynamic responses of such Josephson-junction systems
to external currents.
It is shown that the generalized resistance and the
power spectrum of voltage fluctuations are related via the
fluctuation-dissipation theorem.
Implications of the obtained
relations are also discussed in various experimental situations.
\end{abstract}

\bigskip

\newcommand {\avgl}  {\left\langle\:}
\newcommand {\avgr}  {\:\right\rangle}
\newcommand {\im}    {{\rm Im}}
\newcommand {\re}    {{\rm Re}}


\begin{multicols}{2}

There has been much interest in the dynamics of 
Josephson junctions~\cite{Likhar86} and 
Josephson-junction arrays~\cite{Kim93}, e.g.,
current-voltage characteristics, dynamic resistivity, and
voltage fluctuations.  
Among these, the voltage fluctuations provide direct information 
about the dynamic correlations 
in equilibrium~\cite{Likhar72,Chungx89},
whereas the resistivity probes the response to external
currents~\cite{Ambega80}. 
The latter is also closely related to the relaxation function,
which describes the relaxation behavior towards the equilibrium
state.  These two probes are therefore complementary to
each other, and one may expect, in view of the general idea of the
fluctuation-dissipation (FD) theorem, that there exists
a FD relation between them.
Nevertheless most existing studies have been devoted either to
the resistivity or to the voltage fluctuations,
and the relation between the two has hardly been investigated.
%
Here we thus make use of the linear response theory to derive
the generalized frequency-dependent resistance, and
examine the relation between the generalized resistance 
and the power spectrum of the voltage fluctuations in Josephson-junction
systems.

There are three energy scales in a Josephson-junction system:
the Josephson coupling energy $E_J\equiv\hbar{}I_J/2|e|$, 
the self-charging energy $E_0\equiv{}e^2/2C_0$,
and the junction-charging energy $E_C\equiv{}e^2/2C$, 
where $I_J$ is the Josephson critical current and $C_0$ and $C$ are
the self-capacitance and the junction capacitance, respectively.
In case that the Josephson energy dominates ($E_0,E_C\ll{}E_J$),
the system at finite temperatures can be described by classical
Langevin-type equations of motion for the phases of the order parameter 
on superconducting grains. 
For a single junction, in particular, the equation of motion reads
\begin{equation}
\ddot{\phi} + \gamma\dot{\phi} + \sin\phi
= I + \zeta(t)
  ,
\label{RCSJ:Langevin}
\end{equation}
where $\phi$ is the phase difference between the two grains,
the time has been rescaled in units of the inverse of the Josephson
plasma frequency $\omega_p\equiv \hbar^{-1}\sqrt{8E_CE_J}$, $I$ is the
external direct-current bias in units of $I_J$, and the damping parameter
$\gamma$ is related to the shunt resistance $R$ via
$\gamma\equiv\hbar\omega_p/2|e|RI_J$.  The noise current
$\zeta(t)$ (in units of $I_J$) is characterized by zero
mean and correlation
$\avgl\zeta(t)\zeta(t')\avgr=2\gamma{T}\delta(t{-}t')$, where $T$ is the
temperature in units of $E_J$. 

We introduce the probability distribution function $P(\phi,v,t)$
and write the Fokker-Planck (FP) equation~\cite{Risken89}, 
which corresponds to the Langevin equation (\ref{RCSJ:Langevin}):
\begin{equation}
\frac{\partial}{\partial t}P(\phi,v,t)
= L_{FP}(\phi,v) P(\phi,v,t)
  ,
\label{RCSJ:FP}
\end{equation}
where $v\equiv\dot\phi$ is the voltage (in units of
$\hbar\omega_p/2|e|$) across the junction, and the FP
operator $L_{FP}$ is defined to be
\begin{equation}
L_{FP}(\phi,v)
\equiv \left[
    -\frac{\partial}{\partial \phi}v
    + \frac{\partial}{\partial v}\left(
      \gamma{v} + \sin\phi - I + \gamma{T}\frac{\partial}{\partial v}
    \right)
  \right]
  .
\label{RCSJ:FPop}
\end{equation}
Suppose that the system 
is disturbed by a time-dependent external
current $\delta I(t)$ (in units of $I_J$),
which gives the additional term
$-\delta I(t)(\partial/\partial v)$
in the FP operator.
The resulting change in the average value of $v$ takes the form
\begin{equation}
\delta\!\avgl\!v(t)\!\avgr
=\int_{-\infty}^\infty{dt'}\;
  G(t{-}t')\, \delta I(t')
  ,
\end{equation}
where the linear response function $G(t)$ is given by~\cite{Risken89}
\begin{equation}
G(t)
= -\theta(t)\int_0^{2\pi}d\phi\int_{-\infty}^\infty{dv}\;
  v\,\exp(tL_{FP})\, \frac{\partial}{\partial v}P_{\rm eq}(\phi,v)
  \label{LRF}
\end{equation}
with the equilibrium probability distribution $P_{\rm eq}(\phi,v)$.
Note that $G(t)$ describes the voltage response of the system to
the external current and can be expressed in terms of the correlation function
\begin{eqnarray}
G(t)
& = & \theta(t)T^{-1}
  \avgl v(t)[v(0){-}\alpha(\phi(0),v(0))]\avgr 
  \nonumber \\
& \equiv & \theta(t) T^{-1} \tilde{C}_v(t),
\label{FDT}
\end{eqnarray}
where
\begin{equation}
\alpha(\phi,v)
\equiv \frac{1}{P_{\rm{eq}}(\phi,v)}
  \left( v+T\frac{\partial}{\partial v} \right)
  P_{\rm{eq}}(\phi,v)
  .
\label{dcV-like}
\end{equation}
In the frequency space, Eq.~(\ref{FDT}) takes the simple form
\begin{equation}
2T\,\re\chi(\omega)
= S_v(\omega)
  ,
\label{FDT:w}
\end{equation}
where the generalized resistance $\chi(\omega)$ and 
the voltage power spectrum $S_v(\omega)$ are defined to be the
Fourier transforms of $G(t)$ and $\tilde{C}_v(t)$, respectively.
Note also that it is the real part of $\chi(\omega)$ 
(rather than the imaginary part)
which characterizes the dissipation across the junction.
Equation (\ref{FDT:w}) thus comprises the FD relation in the resistively
and capacitively shunted junction (RCSJ) system,
connecting the generalized resistance (i.e., dissipation)
with the voltage correlation function describing equilibrium fluctuations.

The above relation is to be compared with the standard FD 
theorem~\cite{Bedeau71}, which is applicable to a Hamiltonian system.
Namely, when one can explicitly write
$P_{\rm{eq}}= Z^{-1} e^{-H_{\rm{}eff}/T}$ with a
temperature-independent effective Hamiltonian $H_{\rm{}eff}$,
$\alpha(\phi,v)$ in Eq.~(\ref{dcV-like}) simply reduces to a constant,
giving the standard FD relation. 
In the RCSJ system, however, the periodicity in
$\phi$~\cite{end_note:1} urges the system to have a non-zero probability 
current even in the stationary state.  As a result, the stationary
state cannot be described by an effective Hamiltonian,
and in general $\alpha(\phi,v)$
becomes a dynamical variable depending on $\phi$ and $v$.
Similar features have been pointed out
in the system without the $\ddot{\phi}$ term~\cite{Feigelman},
where the FD relation between the correlation and response of $\phi$
(rather than $\dot{\phi}$) has been considered.
Equation (\ref{FDT:w}), in contrast, concerns the voltage $v\equiv \dot{\phi}$,
which is a physical quantity.
Here it is easy to show that the eigenfunction expansion of 
$P_{\rm{}eq}$~\cite{Risken89}
leads to the identity
$\int{dv}\;(\partial/\partial{}v)P_{\rm{}eq}(\phi,v)=0$, which implies
that the (equilibrium) average of $\alpha(\phi,v)$ is just the average voltage:
$\avgl\alpha(\phi,v)\avgr
= \avgl\!v\!\avgr \equiv \bar{v}$.
It then follows that
at long time scales the correlation
function in Eq.~(\ref{FDT}) reduces to the standard voltage correlation 
function:
$
\tilde{C}_v(t) \approx C_v(t)
\equiv \avgl [v(t){-}\bar{v}][v(0){-}\bar{v}]\avgr
$,
which recovers the standard FD theorem with the voltage power spectrum 
$S_v(\omega)=\int_{-\infty}^\infty dt\;e^{i\omega t}\,C_v(t)$.

On the other hand, in systems with the charging energy dominant over 
the Josephson coupling energy, quantum fluctuation effects should be
taken into account, especially at very low temperatures.  Such a macroscopic 
quantum system can be conveniently described by the quantum phase model:
\begin{equation}
H_0
= 4E_C (n + q)^2 - E_J\cos\phi
  ,
  \label{QPM:SJJ}
\end{equation}
where the number $n$ of excess Cooper pairs and the phase difference
$\phi$ are quantum-mechanically conjugate variables (\,$[n,\phi]=i$\,), 
and $q$ is the external gate charge in units of $2e$.  
For simplicity, we assume that dissipation due to, e.g., quasiparticle
tunneling is negligible in this low-temperature regime, 
and consider the system without the bias current,
where $q$ does not change with time.
A disturbing current $\delta I(t)$ applied to the system leads to
the perturbation Hamiltonian 
$
H_1
= 8E_C (n+q)\,\delta q(t)
  ,
$
where $\delta q(t)\equiv -K\int^t dt'\;\delta I(t')$ with
$K\equiv\sqrt{E_J/8E_C}$.  The standard quantum theory of linear
response then gives the induced voltage across the system in the form
\begin{equation}
\delta\!\avgl\!v(\omega)\!\avgr
= \chi_q (\omega)\, \delta q(\omega)
= \chi (\omega)\, \delta I(\omega)
  ,
\end{equation}
where $\chi (\omega) \equiv (iK/\omega) \chi_q(\omega)$
and $\chi_q (\omega)$ is the Fourier transform of the 
retarded Green's function
$
G^R(t-t') \equiv i\theta(t-t')\avgl [v(t), v(t')]\avgr.
$
The corresponding FD theorem reads
\begin{equation}
(1-e^{-\omega/KT})S_v(\omega)
= 2\,\im{}\chi_q (\omega) = 2(\omega/K)\,\re\chi(\omega),
  \label{QFDT:w}
\end{equation}
where the power spectrum $S_v(\omega)$ is again given by the Fourier transform
of the voltage correlation function $C_v(t)$,
and $\omega$ should be understood as 
$\lim_{\epsilon\to 0^+}(\omega+i\epsilon)$.  
It is pleasing that in the classical limit ($\omega/T \to 0$), 
Eq.~(\ref{QFDT:w}) reproduces the classical
relation Eq.~(\ref{FDT:w}). 
Here without the bias current, we have $\bar{v}=0$ and 
$C_v(t)= \avgl v(t) v(0)\avgr$.
In the presence of the bias current,
$n_0$ increases with time and the unperturbed Hamiltonian $H_0$
depends explicitly on time, which in general does not allow 
the standard derivation. 
Nevertheless when the bias current is sufficiently small, 
it may be incorporated into the disturbing current;
this leads to the same FD relation as that shown in Eq.~(\ref{QFDT:w}).

We now investigate the physical implications of the relations in
Eqs.~(\ref{FDT:w}) and (\ref{QFDT:w}) to several cases.  
First, we consider the case that the external bias current
is smaller than the Josephson critical current ($I<1$). 
In view of Eq.~(\ref{FDT:w}), of particular interest in this case is
the underdamped ($\gamma\ll1$) classical junctions at low temperatures.  
At zero temperature the phase of the (unperturbed) junction
stays at one of the local minima
$\phi = \sin^{-1}\!I$ (mod $2\pi$)
of the wash-board potential
$U(\phi)= -\cos\phi - I\phi$.
Small perturbations then induce the well-known plasma oscillation
in the vicinity of the local minimum,
with the oscillation frequency $\omega_{P}$ given by 
$(1-I^2)^{1/4}$ at $T=\gamma=0$.
At finite but sufficiently low temperatures ($T\ll 1$), 
the noise current stirs
up the small fluctuations around the local minimum, and 
the power spectrum of the voltage is known to become
$S_v(\omega)
= 2\gamma{}T \omega^2 [(\omega^2-\omega_{P}^2)^2 + \gamma^2\omega^2]^{-1}$
~\cite{Likhar86}.
The FD relation in Eq.~(\ref{FDT:w}) then gives
\begin{equation}
\re\chi(\omega)
= \frac{\gamma \omega^2}{(\omega^2-\omega_{P}^2)^2 + \gamma^2\omega^2}
  ,
  \label{FDT:PO}
\end{equation}
which reveals that at low temperatures the generalized resistance 
is temperature-independent.
Note also the behavior $\re\chi(\omega)\to0$ as $\omega\to0$,
which is nothing but the manifestation of the superconducting channel.
A system with strong quantum fluctuations displays
another interesting phenomenon, the Bloch
oscillation~\cite{Likhar85}, at sufficiently low
temperature.  In this case, Eq.~(\ref{QFDT:w}) indicates that
like the voltage spectrum, the generalized resistance should also exhibit 
resonance peaks at the Bloch oscillation frequencies proportional
to the bias current.  

We next consider the Josephson oscillation in an overdamped ($\gamma\gg{}1$) 
Josephson junction
with a large bias current ($I>1$)~\cite{Ambega69}, for which 
the Fokker-Planck equation (\ref{RCSJ:FP}) reduces to~\cite{Risken89}
\begin{equation}
\gamma \frac{\partial}{\partial t}P(\phi,t)
= \frac{\partial}{\partial \phi}\left(
    \sin\phi - I + T\frac{\partial}{\partial \phi}
  \right) P(\phi,t)
  .
\label{OD:FP}
\end{equation}
At zero temperature, the voltage $v(t)$ across such a resistively shunted 
junction (RSJ) displays the Josephson oscillation
with the frequency $\omega_{J}=\gamma^{-1}\sqrt{I^2-1}$, 
which leads to the dc voltage $\bar{v}=\omega_{J}$.
As the temperature is increased from zero, 
the dc voltage also grows, approaching the
Ohmic characteristics.  At the same time, thermal fluctuations
introduce decaying behavior of the probability in addition to
the oscillatory behavior.  The eigenfunction expansion of the 
transition probability~\cite{Risken89} at large time scales 
shows that the leading contribution comes from the lowest eigenvalue, giving
\begin{equation}
P(\phi,t|\phi_0,0)
\sim e^{-\gamma_{J}|t|}\cos(\omega_{J}t)
  + \mbox{higher harmonics}
  ,
\label{TransP}
\end{equation}
where the line width $\gamma_{J}$ of the
Josephson oscillation are related to the dc voltage $\bar{v}(I,T)$ via
$\gamma_{J} = \pi\gamma (d\bar{v}/dI)^2{}T$~\cite{Likhar86}.
While at $T=0$, we obviously have
$\gamma_{J}=0$ and $\omega_{J}=\gamma^{-1}\sqrt{I^2-1}$,
at finite temperatures, they may be estimated numerically to give
$\gamma_{J}\approx \bar{v}=\omega_{J}$ at $T\approx 1$
(see also Sec.~11.3 of Ref.~\cite{Risken89}). 
Equation (\ref{TransP}) leads to the power spectrum
\begin{eqnarray}
S_v(\omega)
& \sim & \frac{\gamma_{J}}{\gamma_{J}^2+(\omega-\omega_{J})^2}
  + \frac{\gamma_{J}}{\gamma_{J}^2+(\omega+\omega_{J})^2} 
  \nonumber \\
& & \mbox{} + \mbox{higher harmonics}
  ,
\label{VNS.apprx}
\end{eqnarray}
which shows correlation peaks at $\omega=\pm \omega_{J}$;
this is followed by resonances of the resistance via Eq.~(\ref{FDT:w}).
It is also of interest to note the crossover behavior
to the high-temperature (noise-dominant) regime at $T\approx 1$,
where the correlation peak in $S_v(\omega)$ disappears.

Heretofore we have concentrated on single-junction systems, but
the generalization to an array system is straightforward. 
The classical equation of motion (\ref{RCSJ:Langevin}) is
easily generalized to any array of Josephson junctions,
e.g., a two-dimensional (2D) square $N\times N$ array:
\begin{eqnarray}
\sum_j C_{ij}\ddot\phi_j
  + \gamma\sum_j\Delta_{ij}\dot\phi_j
  + {\sum_{j}}'\sin(\phi_i-\phi_j)
  & & \nonumber \\
= I_i + {\sum_{j}}'\zeta_{ij}(t)
  & &
\label{RCSJ:JJA}
\end{eqnarray}
with the noise current characterized by
$
\avgl\zeta_{ij}(t)\zeta_{kl}(0)\avgr
= 2\gamma{}T\delta(t)
  (\delta_{ik}\delta_{jl}-\delta_{il}\delta_{jk})
$,
where $\Delta_{ij}$ is the lattice Laplacian, 
$C_{ij}\equiv (C_0/C)\delta_{ij}+\Delta_{ij}$ is the dimensionless
capacitance matrix, 
and the prime restricts the summation
over the nearest neighbors of site $i$.
$I_i$ is the current fed into site $i=(x,y)$, given by 
$I_i = I(\delta_{x,0}-\delta_{x,N})$; this corresponds to that
along one edge of the array ($x=0$), uniform current $I$ is injected into each
site, while along the opposite edge ($x=N$) 
the same current is extracted from each site. 
Similarly, in the quantum regime, the array is described by the
Hamiltonian:
\begin{eqnarray}
H_0
& = & 4E_C\sum_{ij}(n_i + q_i)C_{ij}^{-1}(n_j + q_j) \nonumber \\
& & \mbox{}
  - E_J\sum_{<ij>}\cos(\phi_i-\phi_j)
\end{eqnarray}
with $[n_i,\phi_j]=i\delta_{ij}$,
which is the obvious generalization of Eq.~(\ref{QPM:SJJ}).
With these, we can use the same procedure as that for a single junction, 
which leads to the conclusion that the FD relations in
Eqs.~(\ref{FDT:w}) and (\ref{QFDT:w}) are applicable to
junction arrays as well~\cite{note2}.

To examine the implications of the FD relations to array systems,
we consider a 2D array at zero (direct) current bias, which
is well known to display the Berezinskii-Kosterlitz-Thouless (BKT)
transition~\cite{Berezi71} at $T=T_{BKT}$.
According to the dynamic theory of the BKT
transition~\cite{Ambega80,Ambega78} and ac electrical
measurements~\cite{Hebard80b},
the contribution of the vortex bound pairs is screened out
by the free vortices and the imaginary (inductive) part of the 
frequency-dependent complex impedance sharply decreases to zero
at frequency-dependent temperature $T_\omega$ ($>T_{BKT}$),
which is accompanied by 
a peak of the real (resistive) part.
The FD relation in
Eq.~(\ref{FDT:w}) then suggests that the voltage power spectrum $S_v(\omega)$
as a function of the temperature should also show a peak at $T=T_\omega$. 
The voltage power spectrum, which, to our knowledge, has not been
reported on a 2D array near the BKT transition temperature,
can be measured in equilibrium and complement another equilibrium
measurement obtaining the flux noise spectrum~\cite{Shawxx96}.
In the latter, an interesting relation between the flux noise
spectrum and the frequency-dependent conductivity has been
examined at $T\approx {}T_{BKT}$~\cite{Houlri94}.
It would thus be an interesting topic in the future work to investigate
in this region the connection
between the voltage spectrum and the flux noise spectrum.

For large bias currents ($I>1$), the voltage spectrum $S_v(\omega)$ has
been studied numerically in 2D RSJ arrays~\cite{Chungx89}, 
which indeed has revealed the correlation peaks as in Eq.~(\ref{VNS.apprx}).
Further, it has been observed that the peaks in $S_v(\omega)$ 
disappear at temperature $T\approx 1\, (\equiv E_J/k_B$), which
happens to correspond to the BKT vortex unbinding transition 
temperature in the absence of the bias current.  
Here we point out that the apparent
disappearance of the peak has little to do with the phase transition;
rather Eq.~(\ref{VNS.apprx}) shows that it is just a crossover.  
This is consistent with the result that there
should be no phase transition in the 2D array driven by external currents 
larger than the junction critical current~\cite{MYChoi91a}.
%

This work was supported in part by the BSRI Program, 
Ministry of Education and by the KOSEF through the SRC Program.


\begin{thebibliography}{10}
\bibitem{end_note:1}
Noticed that the Fokker-Plank equation given by
  Eqs.~(\ref{RCSJ:FP}) and (\ref{RCSJ:FPop}) 
  is invariant under the translation in $\phi$ by $2\pi$.
\bibitem{note2}
Here the voltage power spectrum and the
resistive response are measured over the whole system,
giving the frequency dependence in Eqs.~(\ref{FDT:w}) and (\ref{QFDT:w}) 
with no additional spatial dependence.
\end{thebibliography}

\end{multicols}

\end{document}